\documentclass[prl,twocolumn,a4paper,superscriptaddress]{revtex4}

\usepackage{amssymb}
\usepackage{bm}
\usepackage{epsfig}
\usepackage{graphicx}
\usepackage{amsmath}

\def\Ef{E$_{\textrm{\tiny F}}$}

\def\vs{V$_\textrm{s}$}
\def\It{I$_\textrm{t}$}

\def\me{m$_e$}

\def\Gab{\textrm{$\overline{\Gamma}$}}
\def\Xb{$\overline{\textrm{X}}$}
\def\Mb{$\overline{\textrm{M}}$}
\def\GaX{$\overline{\Gamma \textrm{X}}$}
\def\XM{$\overline{\textrm{XM}}$}

\begin{document}

\title{Formation of dispersive hybrid bands at an organic-metal interface}

\author{N. Gonz\'alez-Lakunza$^1$, I. Fern\'andez-Torrente$^2$, K. J. Franke$^2$,
N. Lorente$^3$, A. Arnau$^{1,4}$, and J. I. Pascual$^2$}
\affiliation{$^1$ Departamento de F\'{\i}sica de Materiales UPV/EHU,
Facultad de Qu\'{\i}mica, San Sebasti\'an,
Spain\\
$^2$ Institut f\"ur Experimentalphysik, Freie
Universit\"at Berlin, Arnimallee 14, 14195 Berlin, Germany\\
$^3$ Centre d' Investigaci\'o en Nanoci\`encia i Nanotecnologia (CIN2-CSIC),
Campus de la UAB,  Bellaterra, Spain\\
$^4$ Centro de F\'{\i}sica de Materiales, Centro Mixto CSIC-UPV/EHU, San
Sebasti\'an, Spain }

\date{\today}

\begin{abstract}
An electronic band with quasi-one dimensional dispersion is found at
the interface between a monolayer of a charge-transfer complex
(TTF-TCNQ) and a Au(111) surface. Combined local spectroscopy and
numerical calculations show that the band results from a complex
mixing of metal and molecular states. The molecular layer folds the
underlying metal states and mixes with them selectively, through the
TTF component, giving rise to anisotropic hybrid bands. Our results
suggest  that, by tuning the components of such molecular layers,
the dimensionality and dispersion of organic-metal interface states
can be engineered.
\end{abstract}

\maketitle

The use of organic thin films in electronic devices requires the
existence of electronic bands with high conduction properties.
However, due to their weak intermolecular interactions, organic
materials inherently have narrow bands and low electron mobility
\cite{DimiAdvM02}. This is in contrast to inorganic crystals, in
which strong metallic or covalent bonds between atoms cause charge
carriers to be delocalized and to move in wide bands with high
mobility. Organic-inorganic hybrid materials have been proposed as
the ideal framework to merge the high carrier mobility of the latter
with the advantageous properties of organic materials, i.e. their
ease of processing and growth, low-cost, and tunability of
electronic structure by organic synthesis \cite{DimiSci99}. In spite
of being a promising avenue of research, this approach remains
sustained on empirical bases. A conceptual picture at the molecular
scale describing the cross talk between organic and metallic states
and the formation of organic-metal (OM) hybrid bands is still
missing. This can be provided by a detailed look into the properties
of OM interfaces.

The electronic structure of the OM interface  is determined by the
interaction between molecular and metal states. This interaction causes
re-alignment, splitting, and broadening of molecular states as a response
to charge transfer, structural distortions, and hybridization with metal
states \cite{Kahn}. These processes usually mask the molecular thin film
properties and the interface band structure is dominated by the highly
dispersive metal  states \cite{TautzNat06}. On the other hand, molecules
at the OM interface can scatter \cite{WollPRL02} and confine \cite{Pennec}
metal surface states, mix with them \cite{Ferreri}, or even change their
dynamics \cite{ZhuJPCB05,KanazawaJACS07}. Hence, only through a detailed
understanding of the basic principles behind these interactions one can
envisage designing and tuning the properties and, therefore, the
functionality of an OM interface.

\begin{figure}[tb]
\begin{center}
\includegraphics[width=8cm]{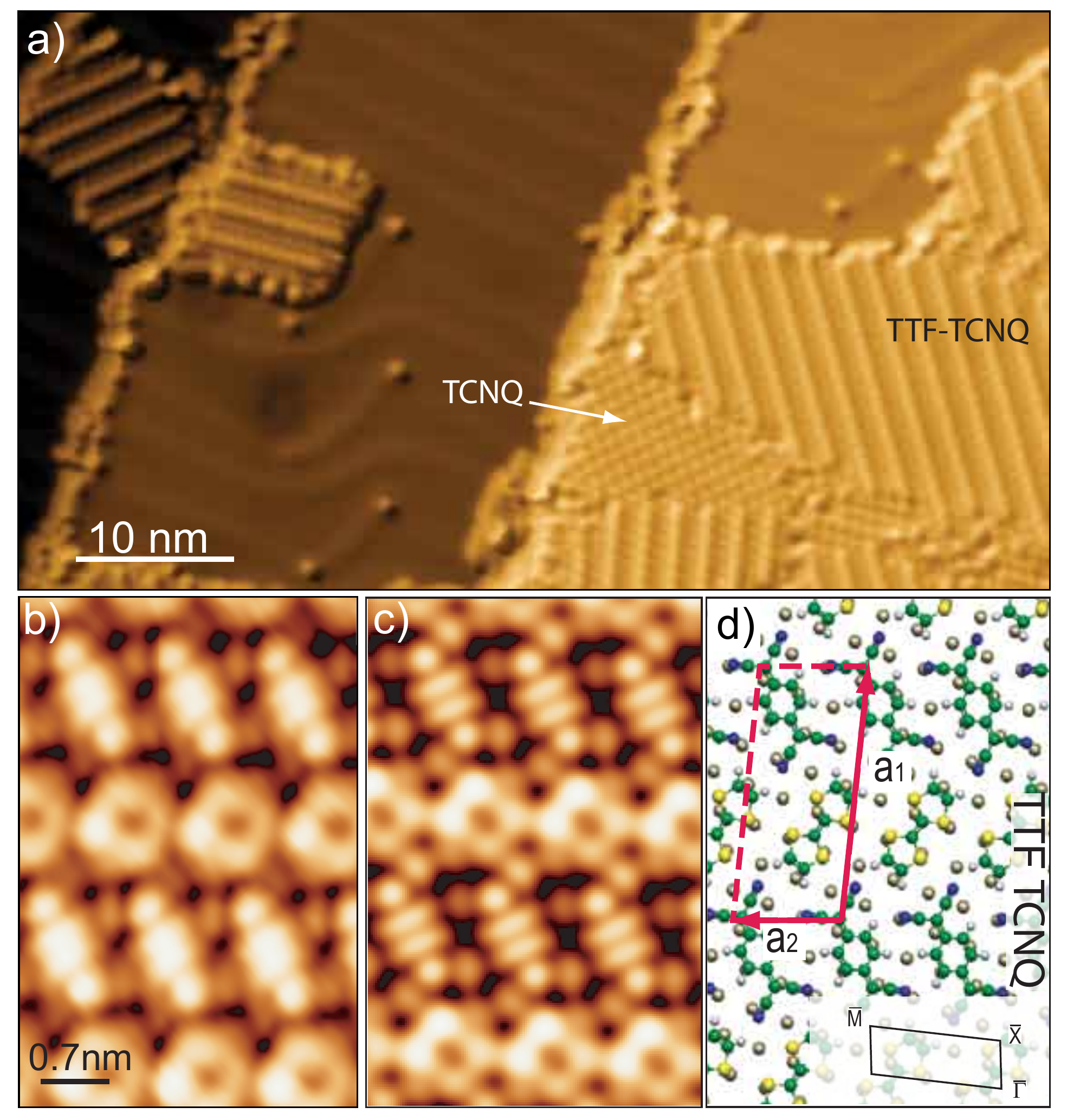}
\end{center}
\caption{(Color online) (a) STM image of TTF-TCNQ domains on
Au(111). Excess TCNQ molecules appear segregated in pure islands
(\vs = 1.7 V; \It=0.1 nA). (b) Intramolecular structure of  TTF and
TCNQ resembling the shape of the HOMO and LUMO, respectively (\vs =
0.3 V; \It=0.4 nA). Data analyzed using WSxM \cite{wsxm}. (c)
Simulated constant current STM image (\vs=1.0 V) using the
Tersoff-Hamann approach \cite{Tersoff} on the DFT minimized
structure shown in (d). C, N, S, H and Au atoms are shown as green,
blue, yellow, white and gold circles, respectively. The vectors a1
and a2 define the surface unit cell (inset shows the corresponding
surface Brillouin zone \cite{note1}). } \label{STM}
\end{figure}

In this Letter, we show that an OM hybrid band is formed at the
interface between the donor-acceptor complex TTF-TCNQ
(tetrathiafulvalene-tetracyanoquinodimethane) and a Au(111)
substrate. The band is characterized by  both a reduced
dimensionality imprinted by the overlayer structure and a large
dispersion reminiscent of its metallic character. By means of a
combined scanning tunneling spectroscopy (STS) and density
functional theory (DFT) study we find that the band originates from
the selective chemisorption of one of the molecular components (TTF)
on the Au(111) surface. Band structure calculations unravel the
complex mixing of metal and molecular states. Our results allow us
to obtain a conceptual understanding about the formation of OM
hybrid bands.

%
%

The experiments were carried out in a custom-built scanning tunneling
microscope (STM), in ultra-high vacuum (UHV) and at a temperature of 5 K.
Both TTF and TCNQ molecules were deposited by UHV sublimation (crucible
temperature 80$^\circ$C) from the solid compound (Aldrich) onto an
atomically clean Au(111) surface held at room temperature. Under these
conditions, TTF and TCNQ self-assemble into mixed domains of alternating
rows of donor and acceptor species with a 1:1 stoichiometry [Fig. 1(a)].
The TTF-TCNQ molecular solid bulk phase is an organic metal with a
one-dimensional-like band structure due to the $\pi$-stacking of each
component and to the transfer of 0.6 electrons from the donor $\pi$-stacks
(TTF) to the acceptor ones (TCNQ) \cite{TTFTCNQreview}. However, on the
Au(111) surface the TTF-TCNQ layer  has different properties because both
TTF and TCNQ lie parallel to the surface, as we can determine from their
intramolecular structure [Fig. 1(b)]. Such adsorption structure is
confirmed by \textit{first principles} calculations using density
functional theory within the generalized gradient approximation
\cite{VASP}. The relaxed structure of a commensurate TTF-TCNQ layer on a
Au(111) surface [Fig. 1(c,d)]  was calculated following the experimental
unit cell \cite{footnote_ss}. We find that the molecular layer is bonded
to the gold surface through TTF via two S-Au bonds per molecule, as in
Ref. \cite{TorrentePRL07}. On the contrary, TCNQ molecules interact weakly
with the metal surface.

\begin{figure}[tb]
\begin{center}
\includegraphics[width=8.5cm]{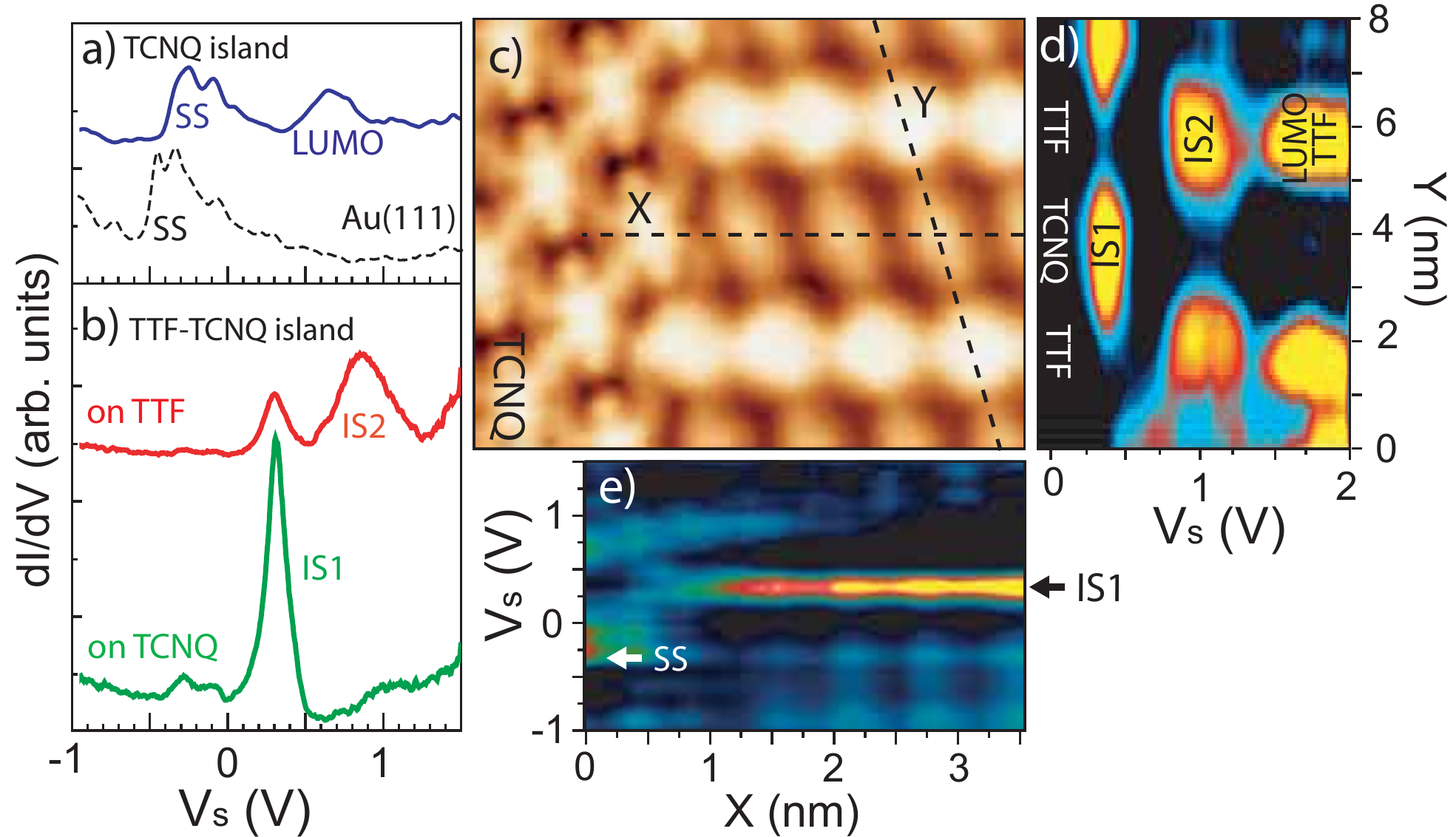}
\end{center}
\caption{(Color online)  Comparison of differential conductance
spectra of (a) TCNQ pure islands (spectrum taken on a nearby clean
Au(111) region shown for comparison) and (b) TTF and TCNQ in mixed
TTF-TCNQ islands. (d,e) The differential conductance spectra is
mapped along the dashed lines shown in the STM image (c)
(\vs=1.9V,\It=0.07nA). The maps clearly show the spatial
localization of IS1 and IS2 features on TCNQ and TTF rows,
respectively.} \label{fig1}
\end{figure}

Scanning tunneling spectroscopy (STS) measurements reveal that the
electronic structure of the TTF-TCNQ/Au(111) interface cannot be
simply understood within a conventional picture of molecular level
alignment. In Fig. 2 we compare the differential conductance (dI/dV)
spectrum of a pure TCNQ island with that of the mixed phase. The
broad LUMO resonance at 0.7 V on pure TCNQ islands [Fig. 2(a)]
contrasts with a sharp peak observed at 0.3 V on TCNQ rows in the
mixed domains [IS1 in Fig. 2(b)].  In addition, the distinctive
onset of the Au(111) surface state (SS) is not observed in the
TTF-TCNQ layer, while another feature (IS2) at 0.8 V appears on the
TTF rows. A more complete picture of the local density of states
(LDOS) of the mixed phase is provided by mapping dI/dV spectra along
the main directions of the  layer [Figs. 2(d,e)]. There, features
IS1 and IS2 appear spatially separated and localized along the TCNQ
and TTF rows, respectively. However, the origin of these peaks
cannot be directly assigned to resonant tunneling through molecular
states, as TTF shows its LUMO resonance at 1.7 V, in agreement with
ref. \cite{TorrentePRL07}. Furthermore, the LUMO peak in the pure
TCNQ island vanishes as we enter in the mixed domain [Fig. 2(e)].
Therefore, these results suggest that the chemisorption of the
TTF-TCNQ layer distorts the electronic structure of the Au(111)
surface, causing additional features to appear in the unoccupied
LDOS.

\begin{figure}[h]
\begin{center}
\includegraphics[width=7cm]{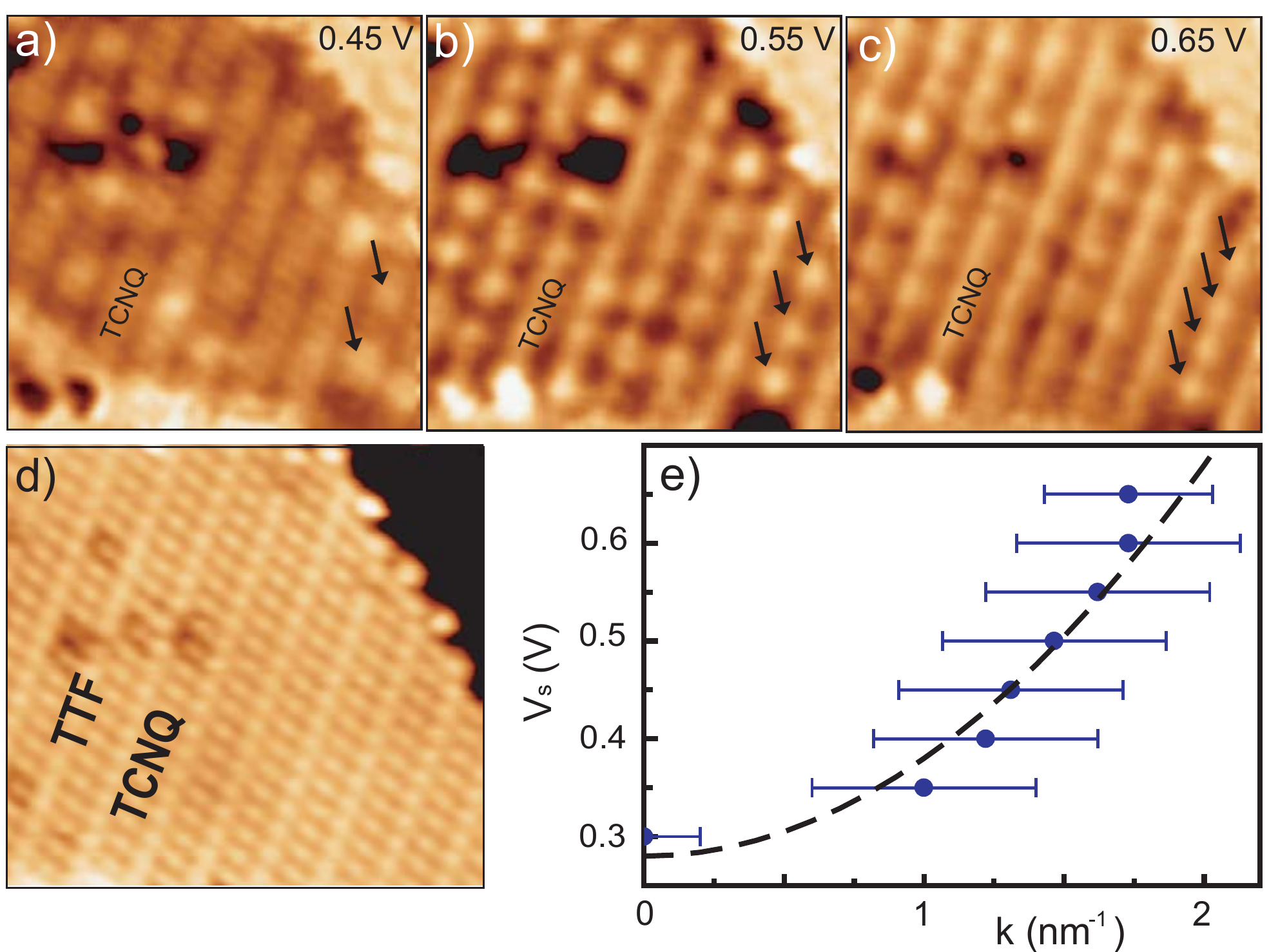}
\end{center}
\caption{(Color online) (a-c) dI/dV maps of a TTF-TCNQ island, shown
in (d), measured at the indicated sample bias values, lying above
the onset of the IS1 resonance (\It=1.0 nA; 16$\times$16 nm$^2$).
(e) Dispersion relation $E(k)$ obtained from data sets similar to
(a-c). The dashed line is a parabolic fit to the data points,
including the IS1 peak position.}
\end{figure}

A hint about the origin of peaks IS1 and IS2 in the spectra is the
dispersive behavior of the former. dI/dV maps taken at sample bias
\vs\ above the IS1 peak exhibit oscillatory patterns centered at the
TCNQ rows [Fig. 3] with a periodicity larger than the intermolecular
distance and changing with \vs. This is a distinctive fingerprint of
(in-plane) quantum interference of electron states with energy e\vs\
\cite{Avouris93}. Furthermore, these oscillations are characteristic
of a quasi-one dimensional electron band, since they are only
observed dispersing along the TCNQ rows. The change in wavelength
with \vs\ follows a parabolic dispersion relation $E(k)=E_0+\hbar^2
k^2/2m^*$ with effective mass $m^{*}=(0.38\pm0.05$)\me\ at $E_0=$
0.28$\pm0.02$ eV above the Fermi energy \Ef\ [Fig. 3(d)]. Electronic
bands of molecular solids exhibit effective masses larger than \me\
\cite{LowDimMetBook,KochPRL06,KollerSci07}. In contrast, the
effective mass  found here is characteristic of metal states,
indicating that the IS1 feature is related to the underlying metal
states \cite{TautzNat06,ZhuJPCB05}.

\begin{figure}[tb]
\begin{center}
\includegraphics[width=7cm]{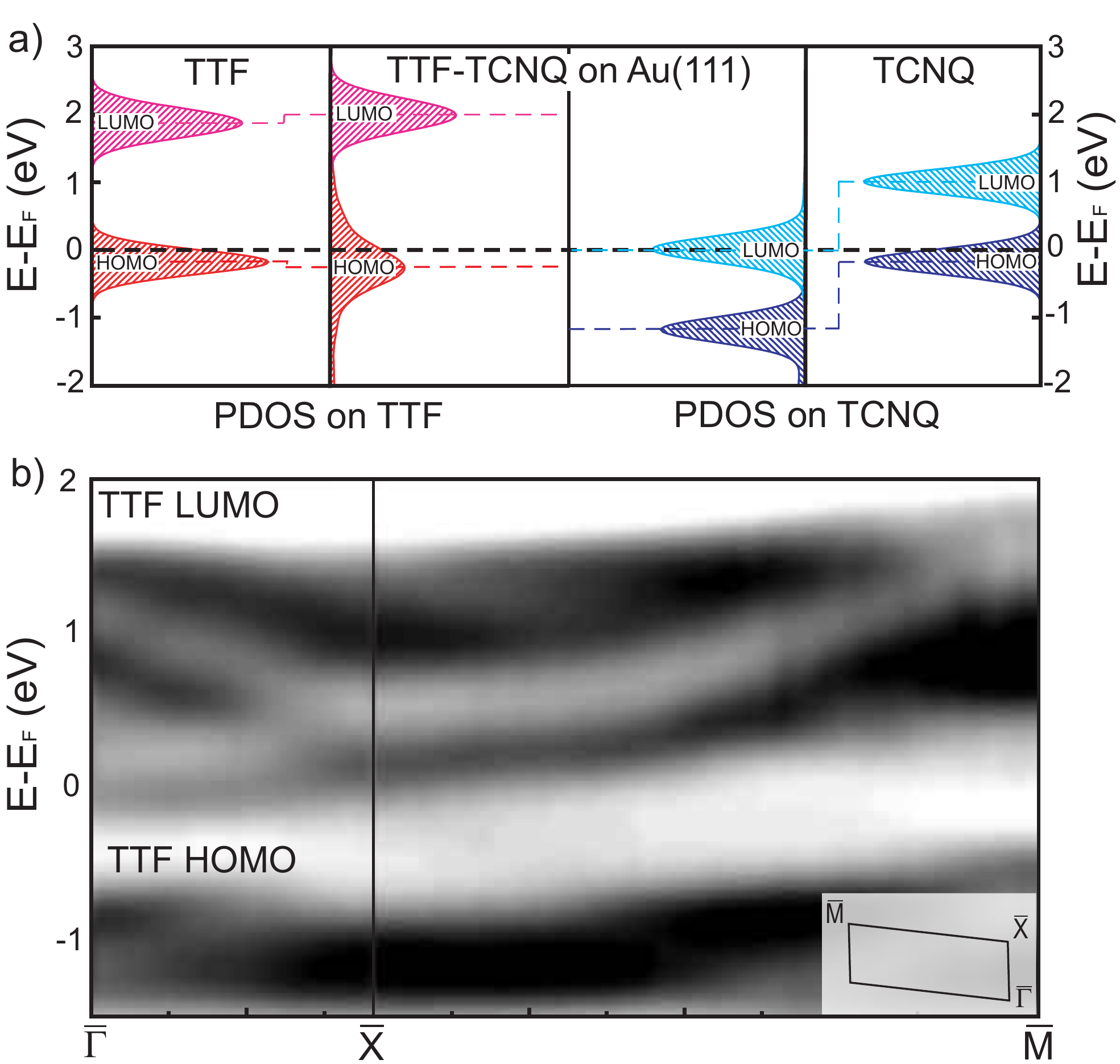}
\end{center}
\caption{(Color online) (a) DOS of the TTF-TCNQ/Au(111) system
projected on TTF and TCNQ orbitals around \Ef, compared with the
PDOS on free molecular chains of each component (leftmost -TTF- and
rightmost -TCNQ- pannels). A gaussian broadening of 250 mV width is
employed. (b) Wave vector resolved PDOS on TTF orbitals along
$\overline{\Gamma \textrm{XM}}$. The PDOS map adds the contribution
of six molecular states around \Ef, since some higher lying
resonances  also show some weight in this energy region. }
\label{teo1}
\end{figure}

To unravel the molecular and metal character of the features IS1 and
IS2 we have calculated the density of states (DOS) of the TTF-TCNQ
layer on Au(111) in the relaxed structure shown in Fig. 1(d). The
role of molecular states is extracted by projecting the DOS (PDOS)
on TTF and TCNQ molecular orbitals [Fig. \ref{teo1}(a)]. We find
that each molecular component maintains its donor/acceptor character
on the metal surface. The TCNQ interaction with the surface is weak,
the PDOS peaks keep their molecular character upon adsorption. Its
LUMO aligns with \Ef , close to the TTF HOMO,
 becoming singly occupied. The TTF HOMO suffers a
notorious broadening as a consequence of its mixing with gold states, and
 becomes partially unoccupied too. Hence, the interaction of the TTF-TCNQ
layer on Au(111) is essentially conducted  
by the TTF molecule.

The PDOS diagrams of Fig. \ref{teo1}(a)  also show that the first
unoccupied molecular state of the TTF-TCNQ film lies at 2.0 eV (TTF
LUMO). The DFT results thus confirm that the origin of the features
IS1 and IS2 in the experimental data cannot be simply ascribed to
the specific molecular orbitals. The PDOS above \Ef\ is rather
dominated by the tail of the TTF HOMO broadened as a consequence of
its hybridization with metal states. The nature of this mixing can
be partially explored by resolving the PDOS vs. electron momentum
along the main directions of the molecular surface Brillouin zone
(SBZ, cf. \cite{note1}). Fig. \ref{teo1}(b) reveals the existence of
two features in the energy region between the TTF HOMO and LUMO
resonances. These features are an evidence of a more complex
substructure hidden behind the broad TTF HOMO resonance.

\begin{figure}[tb]
\begin{center}
\includegraphics[width=8cm]{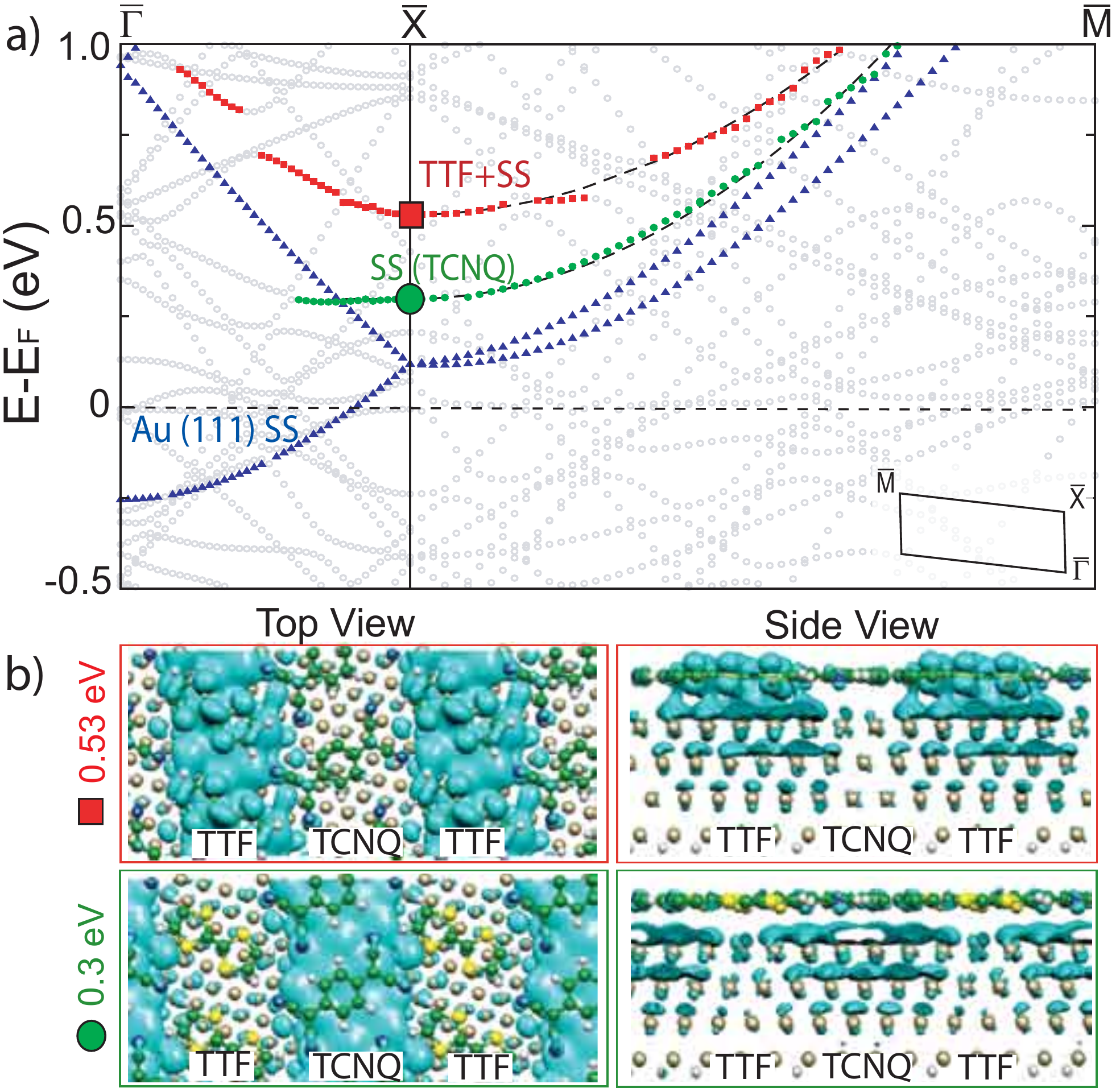}
\end{center}
\caption{(Color online) Calculated electronic structure of the
TTF-TCNQ/Au(111) system. (a) Interface band structure along
$\overline{\Gamma \textrm{XM}}$ directions \cite{ note1}. Green
(\normalsize$\bullet$\small) and red (\tiny$\blacksquare$\small)
dots mark states with charge at the interface. Dashed lines  along
\XM\ show their parabolic fit (m$^*=$ 0.44\me\ and 0.54\me\,
respectively). Mixing with folded bulk states prevents to continue
the green band from \Xb\ to \Gab. The blue bands ($\blacktriangle$)
correspond to the bare Au(111) SS, for comparison. The state is
folded due to the use of a unit cell with the periodicity of the
molecular layer [see Fig. \ref{STM}(d)]. (b) Constant charge density
isosurfaces for two characteristic states located at the \Xb\ point
of the SBZ. }\label{teo2}
\end{figure}

To decipher the role of the metal surface, next we analyze the
calculated band structure along  the \GaX\ and \XM\ directions
\cite{note1}. Two interface bands dispersing away from the \Xb\
point [color points in Fig. \ref{teo2}(a)] are found. The lower
(green) band shows dispersion only along the molecular rows, while
the higher (red) band shows two dimensional anisotropic dispersion
in both directions. The formation of these two interface bands can
be well understood with the help of a simple model \cite{Kittel} of
quasi-free electrons in a weak periodic potential along one surface
direction: (i) the charge re-distribution induced by the organic
layer modulates the potential of the Au(111) surface along the \GaX\
direction with a new periodicity that folds the surface and bulk
bands of the metal, (ii) the coupling with the overlayer
 opens gaps at the SBZ boundary (\Xb\ point) giving rise to the upper
(even) and lower (odd) split hybrid bands (with respect to the TTF rows).

The hybrid character of these bands originates from a mixture of
folded bulk and surface metal states with TTF states, whose strong
influence has already been extracted in Fig. 4(b). The most
interesting evidence of selective mixing of molecular states with
the Au(111) SS is found in partial charge density plots at \Xb. The
lower state (at 0.3 eV) has no molecular character and is located
underneath the TCNQ rows (Fig. 5(b) lower panel). This is in accord
with TCNQ states not contributing to the hybrid band. Furthermore,
its symmetry prevents the lower SS branch from mixing with the TTF
HOMO and thus leaves it essentially as the SS modulated by the new
periodicity imposed by the overlayer. The higher state (at 0.53 eV)
appears centered at the TTF rows and has considerable weight on TTF
(Fig. 5(b) upper panel). As they disperse away from \Ef\, both bands
lose molecular weight and essentially recover the character of the
corresponding SS branches. For the lower band, coupling with the TTF
HOMO and folded bulk metal states introduces broadening away from
the \Xb\ point and closer to the \Gab\ point thus becoming a surface
resonance.

Based on these theoretical results we can now associate the two
observed peaks in the dI/dV spectra with the two calculated
interface OM hybrid bands: IS1 with the lower (green) one and IS2
with the higher (red) one. The calculated partial charge density
maps at \Xb\ [Fig. \ref{teo2}(b)] are in good agreement with the
spatial location of IS1 and IS2 found in the experimental dI/dV
profiles [IS1 on TCNQ rows and IS2 on TTF rows in Fig. 2(d)]. Hence,
the peak structure in the STS spectra is mostly originated from band
states lying close to the \Xb\ point,  because only at this region
of the SBZ they are purely localized at the interface. There, the
lower band is fairly flat across the rows and disperses parallel to
the TCNQ rows in agreement with the one-dimensional dispersion and
sharp line shape of the peak IS1. IS2 appears as a broader peak in
the experimental dI/dV spectra in consistency with the
two-dimensional anisotropic dispersion of the higher band.

In summary, we have shown that the co-adsorption of TTF and TCNQ, an
electron donor and acceptor, on Au(111) gives rise to a mixed phase with a
characteristic interface band structure. While the TCNQ is essentially
unperturbed by the underlying surface, the TTF is hybridized with the
Au(111) surface. The result is the formation of two interface bands with
both molecular and metal character. They exhibit a free-electron metal-like
dispersion and the anisotropic structure of the molecular layer. Our results
suggest that tuning the strength of donor-metal interaction or spacing between
the TTF rows may allow to engineer the organic-inorganic interface band
structure and, hence, the functionality of the molecular thin film.

We thank J. Fraxedas and M. Persson for fruitful discussions and the DFG
for support though Sfb 658. N.G-L. and A.A. thank the Basque Department of
Industry (ETORTEK programme), UPV/EHU, DIPC, and MEC for financial
support. I.F.-T. thanks the Generalitat de Catalunya for her research
grant. N.L. thanks MEC (No. FIS2006-12117-C04-01).


\begin{thebibliography}{99}
\bibitem{footnote_ss}
    The Au(111) surface is described
    using a periodic supercell with a four
    layers Au slab passivated with H atoms \cite{ArnauPRL05} at
    the vacuum metal
    interface opposite to where the adsorbed molecular overlayer sits.
    As shown in Fig. 5 (blue band), in the absence of
    molecular overlayer  the energy position of the surface state at the
    $\bar{\Gamma}$ point is -0.3 eV and the effective mass is m$^{*}$=0.29\me,
    both in good agreement with experimental values
    \cite{ReinertPRB01}.
\bibitem{note1} The surface
    Brillouin zone is defined by the points
    \Gab=(0,0), \Xb=(1/2,0) and \Mb=(1/2,1/2) in units of the reciprocal unit cell vectors b1 and b2.
    \GaX\ and \XM\ correspond to the directions
    across and along the molecular rows, respectively.
\end{thebibliography}
\end{document}